\documentclass[12pt]{article}
\usepackage{graphicx}

\begin{document}

\baselineskip 0.1667in

\begin{center}
{\large \textbf{A Wave Interpretation of the Compton Effect}}

{\large \textbf{As a Further Demonstration of the Postulates of de Broglie}}

\vspace{1cm}

\textsf{Ching-Chuan Su}

Department of Electrical Engineering

National Tsinghua University

Hsinchu, Taiwan

\vspace{1cm}
\end{center}

\noindent \textbf{Abstract}\textit{\ }-- The Compton effect is commonly
cited as a demonstration of the particle feature of light, while the wave
nature of matter has been proposed by de Broglie and demonstrated by
Davisson and Germer with the Bragg diffraction of electron beams. In this
investigation, we present an entirely different interpretation of the
Compton effect based on the postulates of de Broglie and on an interaction
between electromagnetic and matter waves. The speeds of interacting
electrons in the Compton scattering are quite fast and its mechanism relies
heavily on the mass variation. Thus, based on this wave interpretation, the
Compton effect can be viewed as a further demonstration of the postulates of
de Broglie for high-speed particles. In addition to the scattered wave, a
direct radiation depending on the mass variation is predicted, which
provides a means to test the wave interpretation.

\vspace{1.5cm}

\noindent {\large \textbf{1. Introduction}}\\[0.2cm]
In 1923 Compton found that the $x$ rays scattered from free electrons shift
in wavelength. In Compton's explanation of the scattering, the collision
between a photon and an electron in conjunction with the conservation of
energy and momentum is adopted [1]. Thus the Compton effect is commonly
cited as a demonstration of the corpuscular picture of light [2, 3]. On the
other hand, in 1924 de Broglie initiated the concept of wave nature of
matter with the postulate that a particle is associated with a matter wave
of which the frequency and the wavelength are related to the energy and the
momentum of the particle, respectively [3]. This hypothesis of matter wave
led to the introduction of Schr\textrm{\"{o}}dinger's equation, the
Klein-Gordan equation, and the Dirac equation which play the fundamental
role in quantum mechanics. Shortly, in 1927, the matter wavelength was
demonstrated by Davisson and Germer with the Bragg diffraction of electron
beams from a crystal [3]. More recently, various experiments of quantum
interference between matter waves of two coherent beams of electrons,
neutrons, or atoms have been reported to demonstrate the Bragg reflection,
the double-slit diffraction, the gravitational effect, and the Sagnac effect
[4, 5]. Particularly, the effect of earth's rotation has been detected by
the neutron interferometry where the Bragg reflection from slabs of silicon
crystal is used to form a closed path for neutron beams to interfere [5].

In this investigation, based on the postulates of de Broglie, we present an
entirely different interpretation of the Compton effect by dealing with an
interaction between electromagnetic and matter waves. Moreover, it is shown
that the postulates of de Broglie themselves can be derived from the
dispersion of matter wave which in turn is governed by the Klein-Gordan
equation. Under the influence of electromagnetic waves, electrons are
accelerated and a mixed state of matter wave of the interacting electrons is
formed during their state transition. This mixed state leads to a space- and
time-varying medium, from which electromagnetic waves are scattered. Then
the Compton effect corresponds to a constructive interference of
electromagnetic waves which results in dominant scattered waves among
various other scattered waves and direct radiation from the medium. Thereby,
the Compton effect is envisaged as the constructive scattering of
electromagnetic waves from a space- and time-varying medium due to the mixed
state, without explicit resort to the conservation laws for energy and
momentum. Similar scattering mechanisms can also be used to interpret the
Bragg reflection and the Raman scattering from a crystal. The particle
speeds in the Compton effect can be much closer to the speed of light than
those in the aforementioned quantum-interference experiments and its
mechanism relies heavily on the mass variation. Thus, based on this wave
interpretation, the Compton effect can be viewed as a further demonstration
of the postulates of de Broglie for high-speed particles.

\vspace{1cm}

\noindent {\large \textbf{2. Compton Shift}}\\[0.2cm]
When a beam of $x$ rays of a certain wavelength is incident upon a target
made of graphite, the scattered beam shifts in wavelength with a
distribution [1, 2]. For each scattering angle $\varphi $, the scattered
beam tends to have two peaks in the intensity spectrum. The wavelength of
one peak is identical to the incident one, while the second peak shifts to a
longer wavelength. Further, it has been found that this shift depends on the
scattering angle. Quantitatively, the Compton shift in wavelength is given
by [1-3] 
$$
\Delta \lambda /\lambda _{C}=1-\cos \varphi ,\eqno
(1) 
$$
where $\Delta \lambda =\lambda _{s}-\lambda _{i}$, $\lambda _{i}$ and $%
\lambda _{s}$ are the wavelengths of the incident and the scattered beam, $%
\lambda _{C}=h/m_{0}c$ called the Compton wavelength, $m_{0}$ the rest mass
of the electron, and $h$ Planck's constant (see Fig.\hspace{0.1cm}1). It is
seen that the shift increases with the scattering angle to a maximum of $%
2\lambda _{C}$. The Compton wavelength of a free electron is $\lambda
_{C}=2.43\times 10^{-12}$ m. In order for the Compton shift to be
appreciable, the incident wavelength should not be much longer than $\lambda
_{C}$. In Compton's experiment the wavelength of $x$ rays is about 70 pm.

\vspace{0.3cm}

\hspace{2.4cm}\includegraphics[bb=0 0 4in 1.6in]{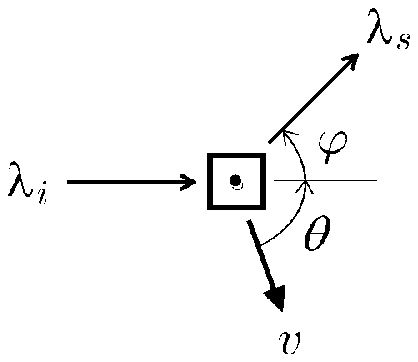}\vspace{0.3cm}

\hspace{0.46cm}%
\parbox{5.5in} {\baselineskip 0.1667in
{\bf Fig.}\hspace{0.1cm}{\bf 1}\hspace{0.2cm}
The Compton effect. The dot represents an electron loosely bound in a target,
which moves with speed $v$ and angle $\theta $ after the collision.

}\vspace{0.6cm}

In order to analyze this shift in wavelength, Compton adopted the collision
between a photon and an electron. By taking $hc/\lambda $ and $h/\lambda $
as the energy and momentum of the photon according to the postulates of
Einstein, the conservation of energy and momentum results in the following
relations 
$$
\left\{ 
\begin{array}{rllc}
hc/\lambda _{i} & \hspace{-0.2cm}=\hspace{-0.2cm} & hc/\lambda
_{s}+m_{0}c^{2}(\gamma -1) & \vspace{3mm} \\ 
h/\lambda _{i} & \hspace{-0.2cm}=\hspace{-0.2cm} & (h/\lambda _{s})\cos
\varphi +\gamma m_{0}v\cos \theta & \vspace{3mm} \\ 
0 & \hspace{-0.2cm}=\hspace{-0.2cm} & (h/\lambda _{s})\sin \varphi -\gamma
m_{0}v\sin \theta , & 
\end{array}
\right. \eqno
(2) 
$$
where $\gamma =1/\sqrt{1-v^{2}/c^{2}}$, $v$ is the speed of the electron
after the collision, and $\theta $ the angle from the recoiling electron to
the incident beam. Then some algebra leads to the Compton shift given by
(1). As the collision between two particles in conjunction with the
conservation of energy and momentum is used, the Compton effect is then
commonly cited as evidence for the corpuscular picture of light.

\vspace{1cm}

\noindent {\large \textbf{3. Klein-Gordan Equation and Postulates of de
Broglie}}\\[0.2cm]
The Klein-Gordan equation proposed to govern the wavefunction $\Psi $ of a
free particle is a nonhomogeneous wave equation given by [6] 
$$
\left\{ \nabla ^{2}-\frac{1}{c^{2}}\frac{\partial ^{2}}{\partial t^{2}}%
\right\} \Psi (\mathbf{r},t)=\left( \frac{m_{0}c}{\hbar }\right) ^{2}\Psi (%
\mathbf{r},t),\eqno
(3) 
$$
where $m_{0}$ is the rest mass of the particle. Suppose that the
wavefunction $\Psi $ is a wave packet composed of plane waves with a narrow
bandwidth. Each component of the plane waves is of the form of a space-time
harmonic like $e^{ikx}e^{-i\omega t}$, where $\omega $ is the angular
frequency and $k$ the propagation constant. Then, for each of the plane
waves, the wave equation reduces to an algebraic equation. That is, 
$$
\omega ^{2}-c^{2}k^{2}=\left( \frac{m_{0}c}{\hbar }\right) ^{2}.\eqno
(4) 
$$
It is seen that the relationship between $\omega $ and $k$ is nonlinear and
hence the matter wave is dispersive.

It is known that the peak of a wave packet moves at its group velocity. Thus
the speed $v$ of a particle can be given by the group speed $v_{g}$ of the
associated wave packet, that is, $v=v_{g}=d\omega /dk$. Then, from the
preceding dispersion relation, one has 
$$
k=\frac{\omega }{c^{2}}v.\eqno
(5) 
$$
On substituting this relation back into the dispersion relation, one
immediately has 
$$
\hbar \omega =\frac{m_{0}c^{2}}{\sqrt{1-v^{2}/c^{2}}}=mc^{2}\eqno
(6) 
$$
and then 
$$
\hbar k=\frac{m_{0}v}{\sqrt{1-v^{2}/c^{2}}}=mv,\eqno
(7) 
$$
where the speed-dependent mass 
$$
m=\frac{m_{0}}{\sqrt{1-v^{2}/c^{2}}}.\eqno
(8) 
$$
The preceding three formulas are just the postulates of de Broglie and the
Lorentz mass variation law. Thus they can be viewed as consequences of the
dispersion of matter wave.

By using a similar wave equation, which is based on the local-ether model of
wave propagation and from which a first-order time evolution equation
similar to Schr\textrm{\"{o}}dinger's equation can be derived and thereby
the particle velocity is evaluated in a quantum-mechanical way as the time
derivative of expectation value of the position operator, the preceding five
formulas have been given alternatively [7, 8]. Meanwhile, one fundamental
difference is that the particle velocity determining the mass, energy, and
momentum is referred to an earth-centered inertial frame for earthbound
phenomena. However, as the linear speed due to earth's rotation is
relatively low, it makes no substantial difference if the particle speed is
referred instead to a geostationary laboratory frame, as done tacitly in
common practice with the Compton scattering.

\vspace{1cm}

\noindent {\large \textbf{4. Wave Interpretation}}\\[0.2cm]
We then go on to present the wave interpretation of the Compton effect,
based on the interaction between electromagnetic and matter waves and on the
scattering of electromagnetic waves. Under the illumination of an
electromagnetic wave, the electrons loosely bound in the target tend to be
accelerated by the incident electric and magnetic fields. Suppose that the
electrons gain a velocity $\mathbf{v}$ from the action. Thus the initial and
the final state of the electrons are respectively of $(m_{0}c^{2},0\mathbf{)}
$ and $(mc^{2},m\mathbf{v)}$ in energy and momentum, which in turn are
represented by the wavefunctions $\Psi _{1}$ and $\Psi _{2}$ that
incorporate the space-time harmonics $e^{-i\omega _{0}t}$ and $e^{-i(\omega
t-\mathbf{k\cdot r})}$, respectively, where $\omega _{0}=m_{0}c^{2}/\hbar $, 
$\omega =mc^{2}/\hbar $, and $\mathbf{k}=m\mathbf{v}/\hbar $. During a
transition from the initial state to the final state, they are expected to
form a mixed state $\Psi =c_{1}(t)\Psi _{1}+c_{2}(t)\Psi _{2}$ with suitable
coefficients $c_{1}$ and $c_{2}$. As the density of electrons is
proportional to the product $\Psi ^{*}\Psi $, this mixed state leads to a
density incorporating a component that varies with space and time in the
form 
$$
e^{i(\Delta \omega t-\Delta \mathbf{k\cdot r})}=e^{i(\omega t-\mathbf{k\cdot
r})}e^{-i\omega _{0}t},\eqno
(9) 
$$
where $\Delta \omega =\omega -\omega _{0}$ and $\Delta \mathbf{k}=\mathbf{k}$%
, and its complex conjugate as well. It is noted that the temporal variation
with $\Delta \omega $ is due to the mass variation. The mixed state results
in time-varying charges and currents, which in turn radiate electromagnetic
waves at the angular frequency $\Delta \omega $. This corresponds to the
celebrated postulate of Bohr for the emission due to state transition of the
electron bound in an excited atom or of high-energy electrons traveling in a
synchrotron. Besides, the mixed state makes the permittivity of a dielectric
medium vary by incorporating the space-time harmonic $e^{i(\Delta \omega
t-\Delta \mathbf{k\cdot r})}$, since the electric susceptibility of a medium
is proportional to the charge density. (The \textrm{complex conjugate} term
has a tendency to make the corresponding scattered wave have a higher
frequency. This inverse Compton scattering [2] is omitted, as the electrons
are initially stationary.)

It is known that the polarization current induced in a dielectric medium is
determined by the product of the susceptibility and the electric field. Thus
the induced charge and current in the space- and time-varying medium tend to
incorporate a key component of which the spatial and temporal variation is
given by the product of the ones of the incident field and the medium. That
is, 
$$
e^{-i(\omega _{s}t-\mathbf{k}_{s}\mathbf{\cdot r})}=e^{-i(\omega _{i}t-%
\mathbf{k}_{i}\mathbf{\cdot r})}e^{i(\Delta \omega t-\Delta \mathbf{k\cdot r}%
)},\eqno
(10) 
$$
where $e^{-i(\omega _{i}t-\mathbf{k}_{i}\mathbf{\cdot r})}$ and $%
e^{-i(\omega _{s}t-\mathbf{k}_{s}\mathbf{\cdot r})}$ denote the space- and
time-variation of the incident wave and of the induced polarization,
respectively. Thereby, $\omega _{s}=\omega _{i}-\Delta \omega $ and $\mathbf{%
k}_{s}=\mathbf{k}_{i}-\Delta \mathbf{k}$. The induced polarization in turn
re-radiates scattered waves at the angular frequency $\omega _{s}$ in any
possible direction. Thus the frequency and hence the wavelength of the
scattered wave are changed, as a consequence of the interaction between
electromagnetic and matter waves. Then we discuss its propagation direction,
as a consequence of the scattering of electromagnetic waves.

A distribution of radiating currents with phase shift given by the
space-harmonic $e^{i\mathbf{k}_{s}\mathbf{\cdot r}}$ behaves like a linear
antenna array with the repeat distance of the antenna element being
vanishing, as far as the radiation pattern is concerned. The progressive
phase shift given by $k_{s}$ determines the pattern. When $k_{s}\leq \omega
_{s}/c$, a main beam with a strong radiation intensity due to constructive
interference can form. Further, $k_{s}$ determines the propagation direction
of the main beam in such a way that the contributions from the various
current elements are in phase along that direction. As in an electronically
steered phased-array radar, every direction is possible. Under the condition 
$$
k_{s}=\omega _{s}/c,\eqno
(11) 
$$
the main beam propagates just in the direction of $\mathbf{k}_{s}$, as in an
end-fire antenna array [9].

Meanwhile, for a current distribution with different $\omega _{s}$ or $%
\mathbf{k}_{s}$ (in direction or magnitude), the scattered wave can also
propagate in the aforementioned direction. However, the radiation is not of
the main beam or the corresponding main beam is much narrower in terms of
angular width, especially for an array much longer than the wavelength [9].
In addition to the space- and time-varying component given by (9), the
permittivity of the target still has a major part which is invariant in
space and time. This component in turn leads to scattered waves of which the
frequency and wavelength remain unchanged from those of the incident wave.
Furthermore, in addition to the re-radiation due to the polarization
current, the mixed state also leads to a direct radiation due to the source
varying with $e^{-i(\Delta \omega t-\Delta \mathbf{k\cdot r})}$, as in the
synchrotron radiation. This radiation has the angular frequency $\Delta
\omega $ and hence the wavelength 
$$
\lambda _{rad}=\lambda _{C}/(\gamma -1)=\lambda _{i}\lambda _{s}/\Delta
\lambda .\eqno
(12) 
$$
However, its intensity does not depend strongly on the scattering angle and
is expected to be low, since $|\Delta \mathbf{k}|\gg \Delta \omega /c$ 
\textrm{as }$\lambda _{i}\gg \lambda _{C}$ and thus the main beam
disappears. This wavelength is inversely proportional to the mass variation
as in the synchrotron radiation, while the Compton shift is approximately
proportional to the variation. It tends to be well separated from $\lambda
_{i}$ and $\lambda _{s}$. Thus the prediction of the direct radiation may
provide a means to demonstrate the mass variation and to test the wave
interpretation.

Aside from the unchanged component of wavelength $\lambda _{i}$, the wave
propagating in the direction of $\mathbf{k}_{s}$ is then mainly the
scattered wave due to the induced current of which the space- and
time-variation is given by $e^{-i(\omega _{s}t-\mathbf{k}_{s}\mathbf{\cdot r}%
)}$, where $\omega _{s}=\omega _{i}-\Delta \omega $ and $\mathbf{k}_{s}=%
\mathbf{k}_{i}-\mathbf{k}$, subject to the phase condition $k_{s}=\omega
_{s}/c$. Thereby, 
$$
k_{i}^{2}+k^{2}-2k_{i}k\cos \theta =(\omega _{i}-\Delta \omega )^{2}/c^{2},%
\eqno
(13) 
$$
where $\theta $ is the angle from $\mathbf{k}$ to $\mathbf{k}_{i}$. The use
of the dispersion relation (4) and a little algebra leads to 
$$
k_{i}k\cos \theta =(\omega _{i}-\omega _{s})(\omega _{i}+\omega _{0})/c^{2}.%
\eqno
(14) 
$$
Then one has 
$$
k_{i}k_{s}\cos \varphi =k_{i}^{2}-(k_{i}-k_{s})(k_{i}+\omega _{0}/c),\eqno
(15) 
$$
where $k_{s}\cos \varphi =k_{i}-k\cos \theta $ is used. It is easy to show
that this relation is identical to (1). Thus the Compton effect can be
interpreted in an entirely different way. Based on the wave interpretation,
the Compton scattering can be viewed as a demonstration of the wave nature
of electrons, of the postulates of de Broglie, and of the wave equation,
instead of a demonstration of the corpuscular picture of light.

In the Bragg diffraction of $x$ rays from a crystal, the Bragg angle of
reflection which corresponds to the constructive interference is determined
by the path difference between waves reflected from two consecutive lattice
planes of a certain spacing [3]. Alternatively, the lattices can be viewed
as a medium of which the permittivity is time-invariant but is space-varying
determined by the lattice constant. Thus the propagation vector of the
reflected wave is changed depending on this constant, as depicted in Fig.%
\hspace{0.1cm}2a, and then the Bragg angle is determined by whether the
phase condition (11) is fulfilled. On the other hand, for an electromagnetic
wave incident upon a time-varying medium, the scattered wave tends to
increase or decrease in frequency. This frequency shift has been observed
experimentally for a microwave propagating in a rapidly growing plasma [10].
Similarly, the Raman effect is associated with the scattering from a sample
where the atoms or molecules are subject to periodic vibration or rotation
[2]. The Raman scattering can also be observed with a crystal where a
lattice wave due to atomic vibration (known as a phonon) is involved [11]. A
sample with a lattice wave can be viewed as a medium of which the
permittivity is space- and time-varying. Thus, as in the Compton scattering,
both the frequency and the propagation vector of the reflected wave are
changed by those of the lattice wave, as depicted in Fig.\hspace{0.1cm}2b.
It is noted that the preceding interpretations are presented without
explicit resort to the conservation laws for energy and momentum. Thus the
conservation of energy and momentum in these phenomena can be viewed as a
consequence of the scattering.

\vspace{0.3cm}

\hspace{0.5cm}\includegraphics[bb=0 0 6in 2.4in]{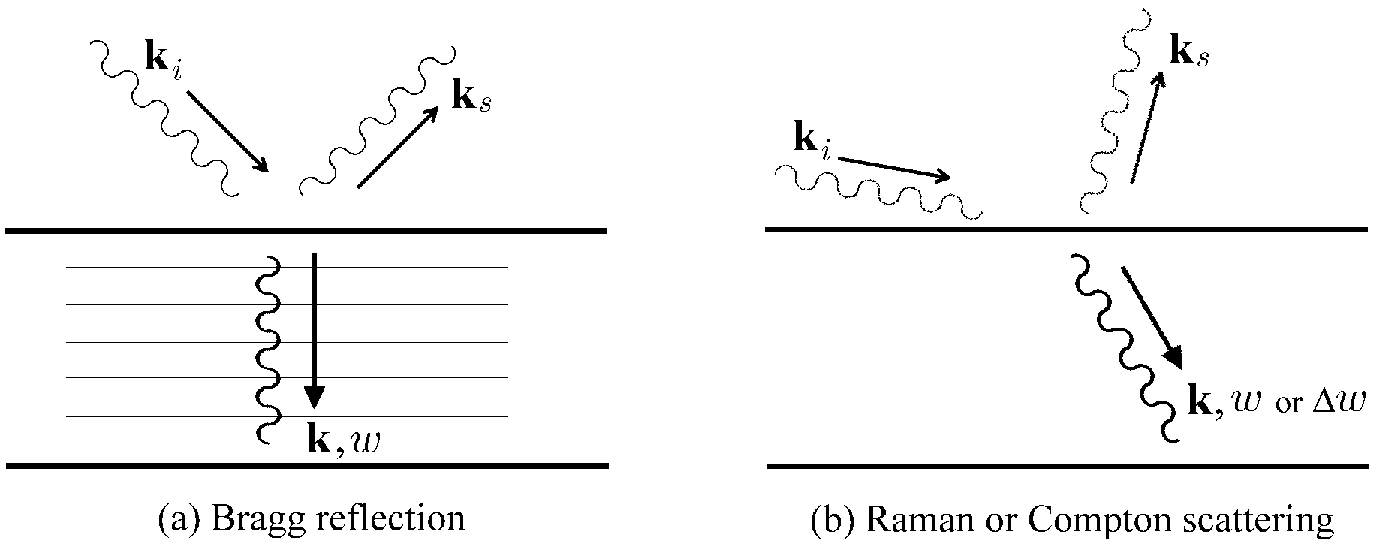}\vspace{0cm}

\hspace{0.46cm}%
\parbox{5.5in} {\baselineskip 0.1667in
{\bf Fig.}\hspace{0.1cm}{\bf 2}\hspace{0.2cm}
Diagrams for the propagation vectors in (a) the Bragg reflection and (b) the Raman or Compton scattering. ${\bf k}_{s}={\bf k}_{i}-{\bf k}$. In the Bragg reflection ${\bf k}$ is determined by the lattice spacing and $\omega =0$; in the Raman scattering ${\bf k}$ and $\omega $ are determined by the lattice wave; and in the Compton scattering ${\bf k}$ and $\Delta \omega $ are determined by the mixed state of matter wave.

}\vspace{0.6cm}

\vspace{1cm}

\noindent {\large \textbf{5. Conclusion}}\\[0.2cm]
Based on the postulates of de Broglie, which in turn are derived from the
dispersion relation for matter wave, an entirely different interpretation is
presented to account for the Compton effect. This approach deals with the
interaction between electromagnetic and matter waves and with the
constructive interference of electromagnetic waves scattered from a space-
and time-varying medium. In this wave interpretation of the Compton
scattering as well as those of the Bragg reflection and the Raman
scattering, the conservation of energy and momentum is not used explicitly
and thus it can be viewed simply as a consequence of the scattering. The
temporal and spatial variation of the target in the Compton scattering in
turn is due to the mixed state of matter wave of interacting electrons
during their state transition and the temporal variation is a direct
consequence of the mass variation. Thus the Compton scattering can be viewed
as a demonstration of the postulates of de Broglie for high-speed particles
and of the wave equation. In addition to the scattered wave, a weak
radiation of which the wavelength is much longer than the one of the
scattered wave is predicted. Its wavelength depends directly on the mass
variation as in the synchrotron radiation. This prediction may provide a
means to test the wave interpretation.

\newpage

\noindent {\large \textbf{References}}

\begin{itemize}
\item[{\lbrack 1]}]  A.H. Compton, ``A quantum theory of the scattering of
x-rays by light elements,'' \textit{Phys. Rev.}, vol. 21, pp. 483--502, May
1923; ``The spectrum of scattered x-rays,'' \textit{Phys. Rev.}, vol. 22,
pp. 409--413, Nov. 1923.

\item[{\lbrack 2]}]  In \textit{McGraw-Hill Encyclopedia of Science }\&%
\textit{\ Technology}, 7th ed. (McGraw-Hill, {New York, }1992), arts.
``Compton effect'', ``Raman effect''.

\item[{\lbrack 3]}]  M. Alonso and E.J. Finn, \textit{Physics} (Wesley, New
York, 1992), chs. 30, 36.

\item[{\lbrack 4]}]  F. Hasselbach and M. Nicklaus, ``Sagnac experiment with
electrons: Observation of the rotational phase shift of electron waves in
vacuum,'' \textit{Phys. Rev. A}, vol. 48, pp. 143--151, July 1993.

\item[{\lbrack 5]}]  J.-L. Staudenmann, S. A. Werner, R. Colella, and A. W.
Overhauser, ``Gravity and inertia in quantum mechanics,'' \textit{Phys. Rev.
A}, vol. 21, pp. 1419--1438, May 1980.

\item[{\lbrack 6]}]  H. Kragh,``Equation with the many fathers. The
Klein-Gordan equation in 1926,'' \textit{Am. J. Phys}., vol. 52, pp.
1024--1033, Nov. 1984.

\item[{\lbrack 7]}]  C.C. Su, {``A local-ether wave equation and
speed-dependent mass and quantum energy,'' }\textit{Eur. Phys. J. B}, vol.
24, pp.\ 231--239, Nov. 2001.

\item[{\lbrack 8]}]  C.C. Su, \textit{Quantum Electromagnetics -- A
Local-Ether Wave Equation Unifying Quantum Mechanics, Electromagnetics, and
Gravitation} (2005),\newline
\texttt{http://qem.ee.nthu.edu.tw}

\item[{\lbrack 9]}]  D.K. Cheng, \textit{Field and Wave Electromagnetics}
(Wesley, New York, 1989), ch. 11.

\item[{\lbrack 10]}]  J. Faith, S.P. Kuo, and J. Huang, ``Frequency
downshifting and trapping of an electromagnetic wave by a rapidly created
spatially periodic plasma,'' \textit{Phys. Rev. E}, vol. 55, pp. 1843--1851,
Feb. 1997.

\item[{\lbrack 11]}]  C. Kittel, \textit{Introduction to Solid State Physics}
(Wiley, New York, 1986), ch. 11.
\end{itemize}

\end{document}